\documentclass{article}
\usepackage{arxiv}
\usepackage[utf8]{inputenc} 
\usepackage[T1]{fontenc}    
\usepackage{hyperref}       
\usepackage{url}            
\usepackage{booktabs}       
\usepackage{amsfonts}       
\usepackage{nicefrac}       
\usepackage{microtype}      
\usepackage{lipsum}
\usepackage{graphicx}
\usepackage{hyperref}       
\usepackage{url}            
\usepackage{booktabs}       
\usepackage{amsfonts}       
\usepackage{amsmath}
\usepackage{color}
\usepackage{color}
\usepackage{braket}
\hypersetup{
    colorlinks=true,
    linkcolor=red,
    filecolor=cyan,      
    urlcolor=magenta,
    citecolor=blue,
}

\title{Approximate evolution for a hybrid system: An optomechanical Jaynes-Cummings model}

\author{L. Medina-Dozal, I. Ramos-Prieto$^{\dagger}$ and J. Récamier$^*$ \\
  Instituto de Ciencias Físicas, Universidad Nacional
Autónoma de México\\ Apdo. Postal 48-3, Cuernavaca, Morelos
62251, México\\
  $^\dagger$\texttt{iranrp123@gmail.com} \\
  $^*$\texttt{pepe@icf.unam.mx} \\
}

\begin{document}
\maketitle

\begin{abstract}
In this work we start from a phenomenological Hamiltonian built from two known systems: the Hamiltonian of a pumped optomechanical system and the Jaynes Cummings Hamiltonian. Using algebraic techniques we construct an approximate time evolution operator $\hat U_{opt}$ for the forced optomechanical system (as a product of exponentials) and take the JC Hamiltonian as an interaction. We transform the later with $\hat U_{opt}$  to obtain a generalized interaction picture Hamiltonian which can be linearized and whose time evolution operator is written in a product form. The analytic results are compared with purely numerical calculations using the full Hamiltonian and the agreement between them is remarkable.
\end{abstract}

\keywords{Jaynes-Cummings \and Optomechanics}
The Jaynes-Cummings model is one of the simplest quantum systems involving the interaction of a single mode quantized electromagnetic field with a two level atom \cite{JC}. Due to its relevance in quantum optics this model has been studied by many authors from the experimental and the theoretical points of view \cite{shore,reviewsJC}. It is well known that the temporal evolution of this system depends upon the properties of the radiation field at the initial time, for instance, if the initial field is a coherent state, the system will present the phenomenon of collapses and revivals for the atomic inversion, this effect has been theoretically and experimentally verified \cite{eberly,rempe,haroche}. Due to its intrinsic relevance in quantum optics, the JC model has been generalized in several forms, for example, assuming that the interaction  between the atom and the field is non linear in the field variables \cite{buck,buzek,cordero,I2014}, the Tavis-Cummings model, where the system is generalized to a group of two level atoms interacting with a one mode field \cite{tavis},  investigating the evolution of the field in presence of a Kerr-like medium \cite{agarwal, werner} or incorporating simultaneously a non linear coupling between the atom and the field and a non linear Kerr-like medium \cite{octavio}.

On the other hand, quantum optomechanics provides a tool to achieve the quantum control of mechanical motion. It does that in devices with mechanical frequencies from a few Hertz to GHz, and masses from $10^{-20}$g to several kilos. It offers a route to determine and control the quantum state of macroscopic objects. Quantum optomechanics provides motion and force detection near the fundamental limit imposed by quantum mechanics \cite{aspelmeyer,kippenberg,meystre}.  To describe the basic physics of cavity optomechanics it is sufficient to consider an optically driven Fabry-P\'erot resonator with one end mirror fixed and the other harmonically bound and allowed to oscillate under the action of radiation pressure from the intracavity light field of frequency $\omega_L$. As  radiation pressure drives the mirror, it modifies the cavity length, and hence the intracavity light field intensity and phase. This results in an optically induced change in the oscillation frequency of the mirror  and optical damping where the optical field acts as a viscous fluid that can damp the mirror motion.   In recent decades, an interest in the motion of mechanical oscillators coupled to oscillation modes in a cavity has resurfaced \cite{corbitt,metzger,thompson}. Some recent applications of this type of resonators include: the LIGO project that uses gravitational wave interferometers whose optical path is modified by radiation pressure \cite{corbitt2}, the cooling of mechanical resonators for the study of the transition between quantum and classical behavior \cite{gigan} and the amplification and measurement of nanometric scale forces \cite{carmon, metzger2}.  

About ten years ago, atom-photon interfaces were proposed as essential building blocks in quantum networks \cite{cirac,duan}. Here, photons are adopted as messengers due to their robustness in preserving quantum information during propagation, while atoms are suited to store the information in stationary nodes. The efficient transfer of quantum information between atoms and photons is essential and requires controlled photon absorption and emission with a very high probability. 
In \cite{hammerer}  the authors achived the strong coupling regime of a mechanical oscillator and a single atom. The coupling between the motion of a membrane (representing the mechanical oscillator) and the atom is mediated by the quantized light field in a laser driven high-finesse cavity. The strong coupling regime provides a quantum interface allowing the coherent transfer of quantum states between the mechanical oscillator and the atoms. Controlled storage of quantum information will require electromagnetically induced transparency (EIT). This technique is widely used to control the absorption of weak light pulses or single photons in atomic ensambles and high finesse cavities. The EIT from a single atom in free space was reported in Ref~\cite{blatt}. There, the authors observed the direct extinction of a weak probe field and electromagnetically induced transparency from a single Barium ion. In \cite{Nori2} the authors studied the transmission of a probe field through a hybrid optomechanical system consisting of a cavity and a mechanical oscillator with a two-level atom. The mechanical resonator is coupled to the cavity field via radiation pressure and to the qubit via the Jaynes Cummings interaction. They find two transparency windows giving rise to an optomechanical analog of two-color EIT.
In contrast to the Fock states, the coherent states are the quantum states whose statistical behavior most resemble the classical one, this has generated considerable interest in using micromirrors for  the generation of coherent mechanical states or even superpositions of them if such micromirrors can be cooled to their quantum ground states \cite{dodonov,zhang,recamier2003,mancini,bose,perez}.
One way to cool a mechanical resonator mode is to use the radiation pressure force exerted by photons in an optical cavity  to damp the Brownian motion of the movable mirror. Since radiation pressure depends on the number of photons present in the cavity, it is to be expected that the cooling of the micro mirror can be manipulated by control of the photon statistics \cite{sumei}.
   
In this work we consider a hybrid optomechanical system composed by a cavity, a mechanical oscillator and a two level atom inside the cavity. The system is pumped by an external laser of frequency $\omega_L$ and amplitude $\Omega$. We construct an approximate time evolution operator for the system and evaluate the temporal evolution of several observables like the number of photons, phonons and the Mandel Q Parameter.

 The paper is organized as follows: In section \ref{Theory} we present the basic theory to obtain a time evolution operator for a tripartite system composed by a forced optomechanical Hamiltonian a one mode cavity and a two level atom inside the cavity (Jaynes-Cummings Hamiltonian). In section \ref{Eval}  we write the observables of interest in a generalized interaction picture and  in section  \ref{Results} we present our numerical results and conclusions.

\section{Theory}\label{Theory}
We begin by considerning a hybrid optomechanical system whose Hamiltonian is
\begin{equation}\label{eq:hamiltonian}
\frac{\hat{H}}{\hbar} =  \frac{\hat H_{0}}{\hbar}+\frac{\omega_a}{2}\hat \sigma_z+\lambda (\hat a \hat \sigma_{+}+ \hat a^{\dagger}\hat \sigma_{-})
\end{equation}
with $\hbar \omega_a$ the energy difference between the ground and the excited atomic states, $\lambda$ the coupling constant between the field and the two level atom and $\hat H_{0}$ describing the simplest pumped optomechanical system given by \cite{Ventura, Law_1995, Vitali_2007,Ghobadi_2001}
\begin{equation}\label{eq:hopt}
\frac{\hat H_{0}}{\hbar}= \frac{\hat H_{opt}}{\hbar} +\Omega \cos(\omega_L t)(\hat a+\hat a^{\dagger}), 
\end{equation}
where
\begin{equation}
\frac{\hat H_{opt}}{\hbar} =\omega_c\hat n+\omega_m \hat N - G \hat n(\hat b+\hat b^{\dagger}).
\end{equation}
Here $\omega_c$, $\omega_m$ are the field and the mechanical oscillator frequencies, $\hat n=\hat a^{\dagger}\hat a$, $\hat N = \hat b^{\dagger}\hat b$ are the number operators for the field and the mechanical oscillator and $G$ is the coupling constant between the field and the mechanical oscillator
given by: $G = \frac{\omega_c}{L}\left(\hbar/2m\omega_m\right)^{1/2}$. $\Omega$ is related to the input laser power, $\omega_L$ is the frequency of the driving field, $\hat a$, ($\hat a^{\dagger}$) are the annihilation  (creation) field operators. The Hamiltonian given by \eqref{eq:hamiltonian}, has already been used in models where {\em hybridization} plays a major role, for example: division of the optical and mechanical fluctuation spectra \cite{Dobrindt}, photon blockade and antibunching \cite{Nori1,Nori2} and in state transfer and entanglement in trapped ions \cite{Biswas}.

We have developed a useful approach to find an approximate time evolution operator for the Hamiltonian $\hat H_0$ when the system does not interact with the environment \cite{optoforzado}. Here we use a similar approach to obtain the time evolution operator of the hybrid  system described by the Hamiltonian  given in \eqref{eq:hamiltonian}. The first thing to take into account is that the time evolution operator associated with $\hat{H}_0$ is given by:
\begin{equation}\label{eq:u0}
\begin{split}
\hat{U}_0(t)&=  e^{\delta +\frac{1}{2}|\beta|^2} e^{\alpha_1 \hat n}e^{\alpha_2 \hat N}e^{(\alpha_3+|\alpha_4|^2/2) \hat n^2}\hat{D}_{\hat{b}}(\alpha_4\hat{n}) \hat D_{\hat a}(\beta),
\end{split}
\end{equation}
where $\hat{D}_{\hat{A}}(\alpha)=e^{\alpha\hat{A}^\dagger-\alpha^*\hat{A}}$ is the Glauber displacement operator (for the derivation of \eqref{eq:u0} see appendix \ref{apA}).  

Once we have obtained the time evolution operator corresponding to the Hamiltonian $\hat H_0$ we transform the interaction to get the approximate interaction Hamiltonian
 \begin{equation}
 \hat H_I^{(1)} = \frac{\hbar\omega_a}{2}\hat \sigma_z +\hbar\lambda\left[ (\hat a +\beta)\hat \sigma_{+}e^{-i\omega_c t}+(\hat a^{\dagger}+\beta^{*})\hat \sigma_{-}e^{i\omega_c t}\right]
 \end{equation}
 where we have maintained  the same level of approximation as the one used to get Eq.~\ref{eq:hiap}. The time evolution operator in the interaction picture satisfies the equation
 \begin{equation}
 i\hbar \frac{\partial \hat U_I^{(1)} }{\partial t}= \hat H_I^{(1)} \hat U_I^{(1)}, \ \ \ \hat U_I^{(1)}(0)=1
 \end{equation}
 It is convenient now to write the interaction Hamiltonian as a the sum of a Hamiltonian containing the set $\{\hat \sigma_{+},\hat \sigma_{-},\hat \sigma_z\}$ and another with $\{ \hat a \hat \sigma_{+}, \hat a^{\dagger}\hat \sigma_{-}\}$.
 \begin{equation}
  \hat H_I^{(1)}= \hat H_{1}^{(1)} +\hat H_2^{(1)}, \ \ \ \ \hat U_I^{(1)} = \hat U_1^{(1)} \hat U_2^{(1)},
\end{equation}
with
\begin{equation}
\begin{split}
 \hat H_{1}^{(1)}& = \frac{\hbar \omega_a}{2}\hat \sigma_z + \hbar\lambda\left( \beta e^{-i\omega_c t}\hat \sigma_{+}+\beta^{*}e^{i\omega_c t}\hat \sigma_{-}\right)\\ 
\hat H_{2}^{(1)}& = \hbar \lambda \left( \hat a \hat \sigma_{+} e^{-i\omega_c t}+\hat a^{\dagger}\hat \sigma_{-}e^{i\omega_c t}\right).
\end{split}
 \end{equation}
 The set of operators in $\hat H_{1}^{(1)}$ is closed under commutation, then its time evolution operator has the form
 \begin{equation}\label{U_1}
 \hat U_1^{(1)} = e^{\alpha_z \hat \sigma_z} e^{\alpha_{+}\hat \sigma_{+}} e^{\alpha_{-}\hat \sigma_{-}}. 
 \end{equation}
While for $\hat U_2^{(1)}$ we have the equation:
 \begin{equation}
 i\hbar \frac{\partial \hat U_2^{(1)}}{\partial t} = \left[ \hat U_1^{(1)\dagger} \hat H_2^{(1)} \hat U_1^{(1)} \right] \hat U_2^{(1)}
 \end{equation}
 transforming the interaction we get:
\begin{equation}
\left[ \hat U_1^{(1)\dagger} \hat H_2^{(1)} \hat U_1^{(1)} \right] \simeq \hbar\lambda\left( \hat a\hat \sigma_{+}e^{-i\omega_c t -2\alpha_z}+\hat a^{\dagger}\hat \sigma_{-}e^{i\omega_c t+2\alpha_z}\right)
\end{equation}
 where we have used the fact that $\lambda \ll \omega_a$. Notice that this interaction Hamiltonian has the form of a Jaynes-Cummings (JC) Hamiltonian, and it is important to highlight that the total number of excitations remains constant. \\ 
 The functions $\alpha_i$ satisfy the equations:
 \begin{equation}
 \begin{split}
 \dot \alpha_z& = -i\left( \frac{\omega_a}{2}-2\lambda \beta^{*}e^{2\alpha_z+i\omega_c t}\alpha_{+}\right) \\
 \dot \alpha_{+} &= -i\lambda \left(\beta e^{-2\alpha_z-i\omega_c t}+2\beta^{*}\alpha_{+}^2 e^{2\alpha_z+i\omega_c t}\right) \\
 \dot \alpha_{-} &= -i\lambda \beta^{*} e^{2\alpha_z+i\omega_c t}.
 \end{split}
\end{equation}
 Now we introduce the operators \cite{cordero,rodriguez,pscr}:
 \begin{equation}
 \hat c = \frac{1}{\sqrt{\hat M}}\hat a \hat \sigma_{+}, \ \ \ \ \hat c^{\dagger}= \hat a^{\dagger}\hat \sigma_{-}\frac{1}{\sqrt{\hat M}} 
 \end{equation}
 with $\hat M = \hat n +\frac{1}{2}(1+\hat \sigma_z)$ the total number of excitations in a given ladder. The basis states for the JC Hamiltonian are $\{ |n,e\rangle, |n+1,g\rangle\}$ ($M=n+1$)  corresponding to a state where the atom is in its excited state and the field has $n$ photons and a state where the atom is in its ground state and the field has $n+1$ photons. The state $|0,g\rangle$ ($M=0$) where the atom is in its ground state and the field in the vacuum state does not couple with any state. The action of these operators upon the basis states is:
 \begin{equation}
\begin{split}
\hat c|n,e\rangle &= 0,\qquad \hat c|n+1,g\rangle = |n,e\rangle\\
\hat c^{\dagger}|n,e\rangle& = |n+1,g\rangle,\qquad \hat c^{\dagger}|n+1,g\rangle = 0\\
\hat M |n,e\rangle& = (n+1)|n,e\rangle,\qquad \hat M |n+1,g\rangle = (n+1)|n+1,g\rangle .
\end{split}
 \end{equation}
From the above expressions we obtain the commutation relations
\begin{equation}
[\hat c, \hat c^{\dagger}] = \hat \sigma_z, \ \ \ \ [\hat \sigma_z,\hat c] = 2\hat c, \ \ \ \ [\hat \sigma_z,\hat c^{\dagger}] = -2\hat c^{\dagger} 
\end{equation}
and $\hat c^2$, $\hat c^{\dagger 2}$ acting upon any basis state is zero. The interaction Hamiltonian can be written in terms of the operators $\hat c$, $\hat c^{\dagger}$ as:
\begin{equation}
\left[ \hat U_1^{(1)\dagger} \hat H_2^{(1)} \hat U_1^{(1)} \right] =\hbar \lambda \sqrt{n+1}\left( \hat c e^{-i\omega_c t-2\alpha_z}+\hat c^{\dagger}e^{i\omega_c t+2\alpha_z}\right)
\end{equation}
then we have
\begin{equation}
i\hbar\frac{\partial \hat U_2^{(1)}}{\partial t}= \hbar \lambda \sqrt{n+1}\left( \hat c e^{-i\omega_c t-2\alpha_z}+\hat c^{\dagger}e^{i\omega_c t+2\alpha_z}\right)\hat U_2^{(1)} 
\end{equation}
whose solution has the form
\begin{equation}\label{U_2}
\hat U_{2}^{(1)} = e^{\epsilon_1\hat c^{\dagger}} e^{\epsilon_{2}\hat c} e^{\epsilon_3 \hat \sigma_z}
\end{equation}
with complex, time dependent functions $\epsilon_i$ such that
\begin{equation}
\begin{split}
 \dot \epsilon_1 & = -i \lambda
 \sqrt{n+1}\left(e^{i\omega_c t+2\alpha_z}-\epsilon_1^2 e^{-i\omega_c t-2\alpha_z}\right) \\
 \dot \epsilon_2 & = -i\lambda \sqrt{n+1}(1+2\epsilon_1 \epsilon_2)e^{-i\omega_c t-2\alpha_z} \\ 
 \dot \epsilon_3 & = -i\lambda \sqrt{n+1}\epsilon_1 e^{-i\omega_c t-2\alpha_z} 
\end{split}
\end{equation}
 and with the initial condition $\epsilon_1(0)=\epsilon_2(0)=\epsilon_3(0)=0$.
 
Finally, taking into account the above relationships and in particular \eqref{eq:u0}, \eqref{U_1} and \eqref{U_2}, the full time evolution operator for the hybrid system is:
 \begin{equation}
 \hat U(t)= \hat U_0(t) \hat U_{1}^{(1)}(t) \hat U_2^{(1)}(t),
 \end{equation}
 where each term has been written as a product of exponentials and can be applied easily to any given initial state so that the construction of the evolved wavefunction is relatively straightforward. This result is our main contribution, it is a major challenge to obtain analytic expressions for the evolution of forced optomechanical systems even when the system is not an open quantum system.
 
\section{Evaluation of observables}\label{Eval}
Let us consider an initial state given by $|\Psi(0)\rangle = |n\rangle\otimes |e\rangle\otimes|\Gamma\rangle$ corresponding to cavity with $n$ photons, a two level atom in its excited state and a mechanical oscillator in a coherent state $\Gamma$. Applying the operator $\hat U_I^{(1)}= \hat U_1^{(1)} \hat U_2^{(1)}$ to the initial state we get:
\begin{equation}
\begin{split}
\hat U_{2}^{(1)}|n,e\rangle \otimes |\Gamma\rangle =& e^{\epsilon_3}\bigg[|n,e\rangle +\epsilon_1 |n+1,g\rangle \bigg] \otimes |\Gamma\rangle\\
\hat U_{1}^{(1)}\left[\hat U_{2}^{(1)}|n,e\rangle \otimes |\Gamma\rangle\right]=& e^{\epsilon_3}\bigg[ e^{\alpha_z}(1+\alpha_{+}\alpha_{-})|n,e\rangle+\alpha_{-}e^{-\alpha_z}|n,g\rangle \bigg] \otimes |\Gamma\rangle\\&+ e^{\epsilon_3}\epsilon_1\bigg[e^{-\alpha_z}|n+1,g\rangle+\alpha_{+}e^{\alpha_z}|n+1,e\rangle \bigg]  \otimes |\Gamma\rangle 
\end{split}
\end{equation}
due to the forcing term in the Hamiltonian the total number of excitations is no longer constant; in contrast with the JC Hamiltonian.
This state can be written as:
\begin{equation}
\begin{split}
\hat U_{I}^{(1)}|n,e\rangle \otimes |\Gamma\rangle =& \bigg[c_1(t)|n,e\rangle +c_2(t)|n,g\rangle +c_3(t)|n+1,g\rangle+c_4(t)|n+1,e\rangle\bigg]\otimes |\Gamma\rangle
\end{split}
\end{equation}
If instead of a number state for the field we have a coherent state $|\alpha\rangle$, we get
\begin{equation}\label{eq:psint}
\begin{split}
|\Psi(t)\rangle_I = &  \hat U_{I}^{(1)}|\alpha,e\rangle \otimes |\Gamma\rangle\\ =&\sum_{n=0}^{\infty} c_n \bigg[c_1(t)|n,e\rangle +c_2(t) |n,g\rangle+  c_3(t)|n+1,g\rangle+c_4(t)|n+1,e\rangle\bigg] \otimes |\Gamma\rangle 
\end{split}
\end{equation}
where $c_n = \exp[-\frac{1}{2}|\alpha|^2] \alpha^n/\sqrt{n!}$.
The time evolution operator $\hat U_0$ does not involve the atomic degrees of freedom, then we can use Eq.~\ref{eq:psint} to evaluate the atomic evolution. For instance, the probability to find the atom in its excited state at time $t$ is given by
\begin{equation}
P_e(\alpha,t) = |\langle e|\Psi(t)\rangle_I|^2 = \left|\sum_{n=0}^{\infty} c_n\left[c_1(t)|n\rangle + c_4(t)|n+1\rangle\right]\right|^2 
\end{equation}
\begin{figure}
\begin{center}
\includegraphics{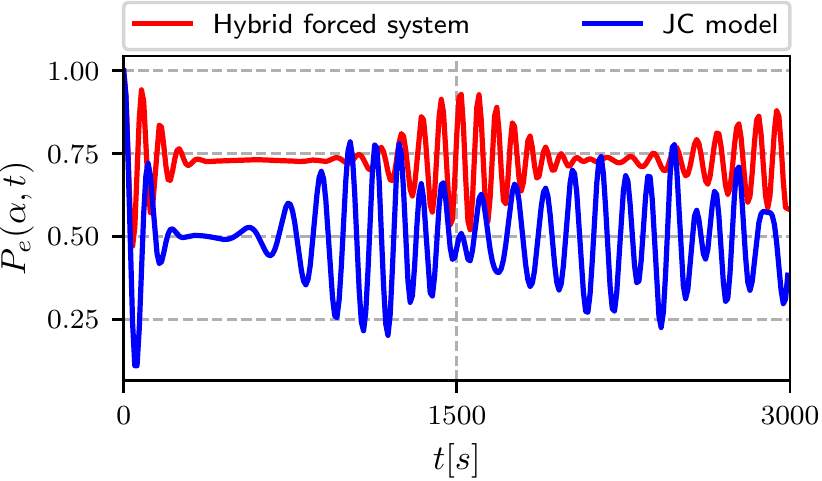} 
\caption{Probability to find the atom in its excited state $P_e(\alpha,t)$ with $\alpha=2$, $\omega_c$=1, $\omega_a=.95 \omega_c$, $\omega_L =0.5\omega_c$, $\lambda=0.0125$, $\Omega = 0.01$. In red we show the case for a forced system, in blue the JC model result.} 
\label{fig_1}
\end{center}
\end{figure}
 In figure \ref{fig_1} we show the probability for the atom to be in its excited state for the hybrid pumped system (red) and for the JC Hmiltonian (blue). The initial state of the cavity is a coherent state with an average number of photons $\bar n = 4$ and atom-cavity coupling constant $\lambda = 0.0125\omega_c$, the pumping amplitude is $\Omega=0.01 \omega_c$ and the atomic frequency is $\omega_a = 0.95 \omega_c$ with the cavity frequency set as $\omega_c =1$. In both cases we can see the usual pattern of quantum collapse and revivals present in the JC model, however the length of the collapse and the definition of the revivals is not the same. In the hybrid pumped case, the time between the collapse and the first revival is longer than in the JC case; the definition of the revival is more definite in the pumped case than in the JC case and the probability to find the atom in its excited state is larger for the pumped case. 

Let us consider now the average value of the photon number operator; it is given by:
\begin{equation}
\begin{split}
\langle \hat n(t) \rangle& = \langle \Psi(t_0)|\hat{U}_I^{(1)\dagger}\hat{U}_0^{\dagger} \hat{n} \hat{U}_0\hat{U}_I^{(1)} |\Psi(t_0)\rangle\\
&= \ _I\langle \Psi(t)|\hat n_I(t) |\Psi(t)\rangle_I
\end{split}
\end{equation}
with $\hat n_I(t)$ the photon number operator in the interaction picture. Taking the explicit form of the operator $\hat U_0$ (see Eq.~\ref{eq:u0}) we obtain
\begin{equation}
\hat n_I(t) = \hat n + \beta^{*} \hat a +\beta \hat a^{\dagger} + |\beta|^2
\end{equation}
and $|\Psi(t)\rangle_I$ given by Eq.~\ref{eq:psint}. For the phonon number operator we get
\begin{equation}
\hat N_I(t) = \hat N + \left(\alpha_4\hat b^{\dagger} + \alpha_4^{*} \hat b\right)\hat n_I(t) +|\alpha_4|^2 \hat n_I^2 (t) 
\end{equation}
and we see that the phonon number operator depends on the number of photons present in the cavity. Since the pumping term modifies the photon number, then it will also modify the phonon number evolution. We can now evaluate observables like the Mandel Q parameter and the photon and phonon dispersions.  We present our numerical results in the following section.

\section{Numerical results, unitary evolution.}\label{Results}
\begin{figure}
\begin{center}
\includegraphics{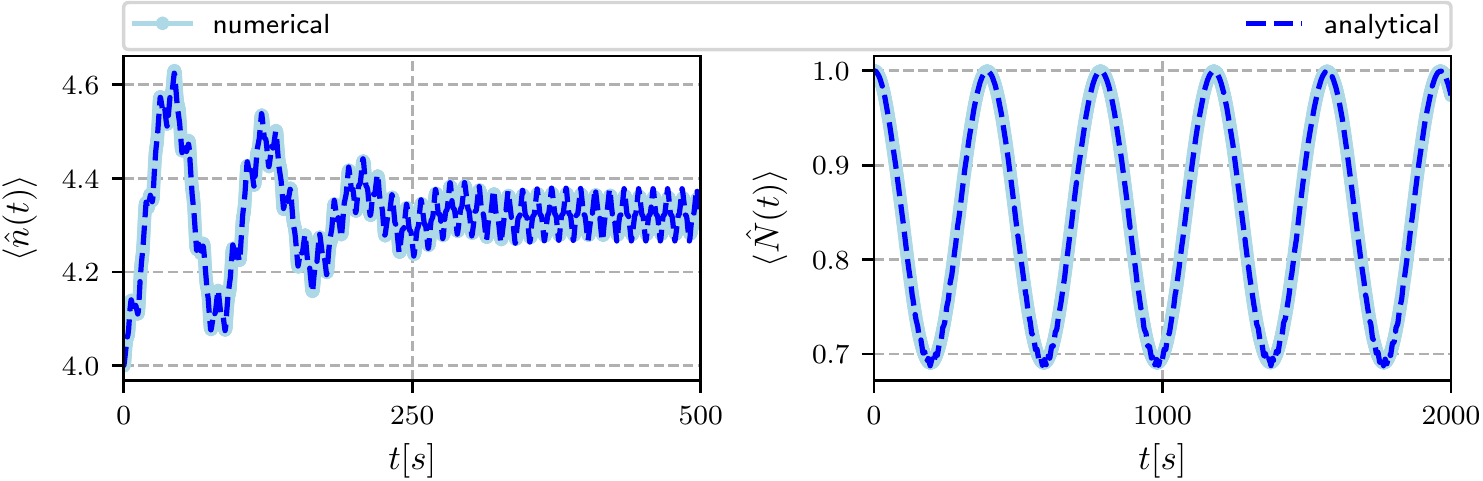} 
\caption{Temporal evolution of the photon number operator (top) and temporal evolution of the phonon number operator (bottom). Analytical results in dark-blue, numerical results in light-blue. Hamiltonian parameters $\alpha=2$, $\omega_c$=1, $\omega_a=.95 \omega_c$, $\omega_L =0.5\omega_c$, $\omega_m = 0.016 \omega_c$, $G=0.00032 \omega_c$, $\lambda=0.0125 \omega_c$, $\Omega = 0.01\omega_c$.} 
\label{figura2}
\end{center}
\end{figure}

In order to test the validity of our approximations, we also made a purely numerical calculation of the average value of the photon, phonon number operators using Python \cite{python}. In figure \ref{figura2} we show the numerical and the analytical results for the temporal evolution of the photon number operator and the phonon number operator for Hamiltonian parameters specified in the caption. The evolution is done for the interval  $0\leq \omega_c t\leq 500$  for the photons and  $0\leq \omega_c t\leq 2000$ for the phonons. We can see an excellent agreement between the analytic and the numerical calculations. For the photons we used an initial coherent state with $\alpha=2$ and for the phonons a coherent state with $\Gamma =1$. Notice that the pumping frequency is far from the resonace cavity frequency $\omega_c$.

\begin{figure}
\begin{center}
\includegraphics{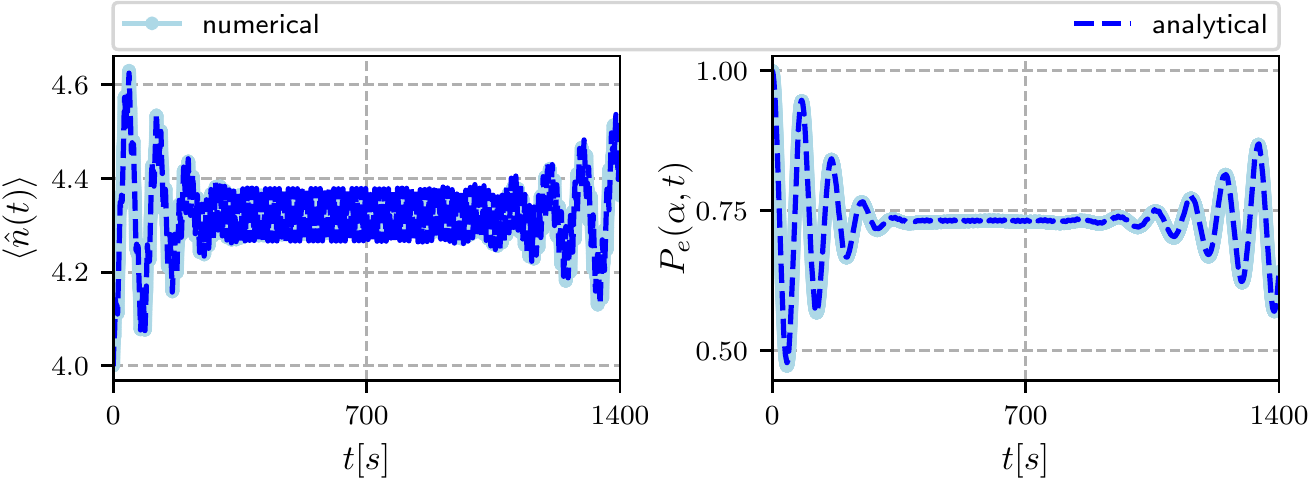} 
\caption{Temporal evolution of the photon number operator (top) and probability to find the atom in its excited state (bottom) with Hamiltonian parameters $\alpha=2$, $\omega_c$=1, $\omega_a=.95 \omega_c$, $\omega_L =0.5\omega_c$, $\omega_m = 0.016 \omega_c$, $G=0.00032 \omega_c$, $\lambda=0.0125 \omega_c$, $\Omega = 0.01\omega_c$. We present numerical (light-blue) and analytical (dark-blue) calculations} 
\label{fig_3}
\end{center}
\end{figure}

In figure \ref{fig_3} we show the temporal evolution of the photon number operator with initial condition $\alpha=2$ corresponding to 
 $\langle \hat n \rangle =|\alpha|^2 = 4$ (top)  and the probability to find the atom in its excited state (bottom). We see an exchange of excitations between the atom and the field, the probability for the atom to remain in its excited state $P_e(\alpha,t)$  decreases to about $0.5$ and at the same time the average number of photons increases to about 4.5, notice also the rapid oscillations with small amplitude around an average value for the number operator, these are due to the forcing term. The overall behavior of the photon number can be guessed from $P_e(\alpha,t)$.

\begin{figure}
\begin{center}
\includegraphics{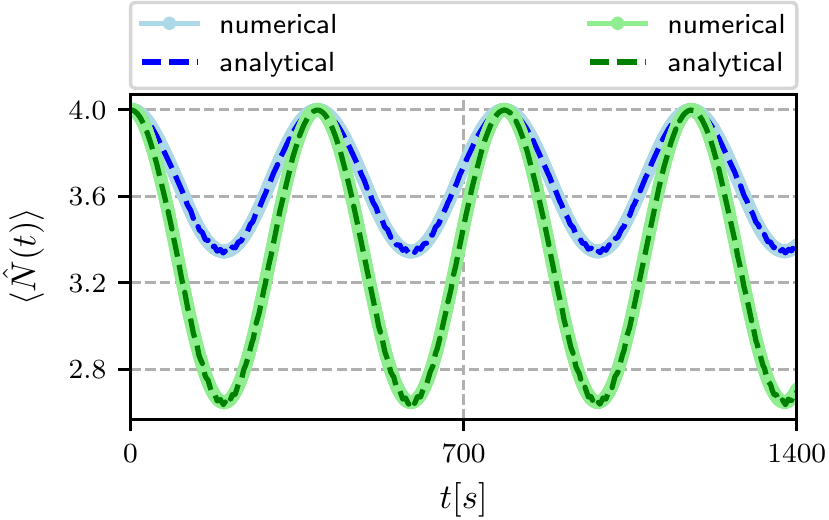}
\caption{Temporal evolution of the phonon number operator with Hamiltonian parameters $\omega_c$=1, $\omega_a=.95 \omega_c$, $\omega_L =0.9\omega_c$, $\omega_m = 0.016 \omega_c$, $G=0.00032 \omega_c$, $\lambda=0.0125 \omega_c$, $\Omega = 0.01 \omega_c$ and  $\left\{\alpha,\Gamma\right\}=\left\{2,2\right\}$ (blue), and $\left\{\alpha,\Gamma\right\}=\left\{3,2\right\}$ (green).} 
\label{fig_4}
\end{center}
\end{figure}

\begin{figure}
\begin{center}
\includegraphics{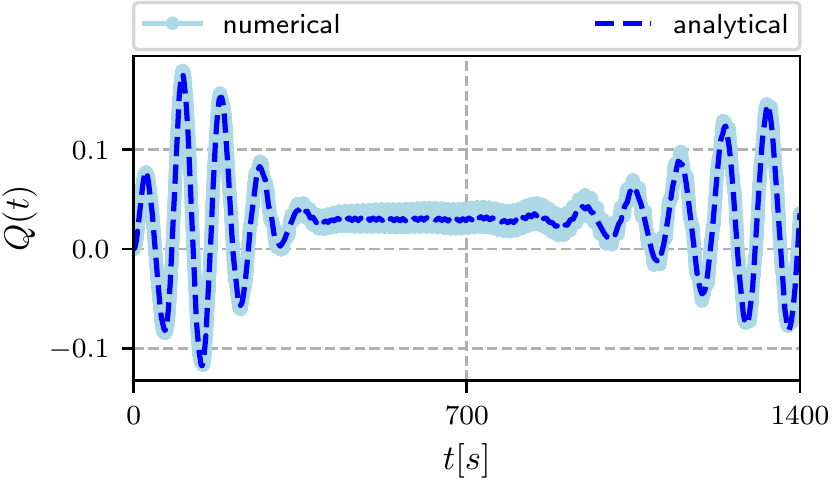}
\caption{Temporal evolution of the Mandel parameter $Q(t)$ with Hamiltonian parameters $\omega_c$=1, $\omega_a=.95 \omega_c$, $\omega_L =0.5\omega_c$, $\omega_m = 0.016 \omega_c$, $G=0.00032 \omega_c$, $\lambda=0.0125 \omega_c$, $\Omega = 0.01 \omega_c$ and  $\alpha=2$.} 
\label{fig_5}
\end{center}
\end{figure}

In figure \ref{fig_4} we show the temporal evolution of the average number of phonons for different amplitudes of the cavty field and Hamiltonian parameters given in the caption. The initial state of the atom is the excited state. In blue we show the case when the initial state of the field is a coherent state with $\alpha=2$ and $\langle \hat n(0)\rangle = |\alpha|^2 = 4$ and  in  green  we plot the case when the initial state of the field is a coherent state with $\alpha = 3$, $\langle \hat n(0)\rangle = 9$. In both cases the initial state of the mechanical oscillator is a coherent state with  $\Gamma= 2$, $\langle \hat N(0)\rangle =4$,
 We have used a pump frequency near resonance $\omega_L=0.9 \omega_c$. Since we are dealing with a red detuning we expect power flow from the mechanical mode to the optical mode \cite{kippenberg-opt} (cooling of the mechanical mode). We see that $\langle N(t)\rangle$ evolves periodically with the frequency of the mechanical oscillator, it decreases from its initial value and after a period it returns to it. Notice that the decrease is larger for the case when the average number of photons is larger so that one can manipulate the number of phonons by means of the interaction time, the amplitude of the cavity field and the frequency of the forcing term. We also show in this plot the results obtained with a purely numerical calculation and we can see a very good agreement between them.
To end this section we present the temporal evolution of the Mandel $Q$ parameter defined as:
\begin{equation}
Q(t) = \frac{\langle \hat n^2(t)\rangle -\langle \hat n(t)\rangle^2}{\langle \hat n(t)\rangle } -1
\end{equation}
for a state with $Q$ in the range $-1\leq Q < 0$ the statistics is sub-Poissonian, and if $Q>0$, super-Poissonian. For a coherent state $Q=0$. In figure \ref{fig_5} we plot the temporal evolution of the $Q$ function for an initial coherent state $| \alpha\rangle$ with $| \alpha |=2$. It starts at zero as corresponds to a coherent state, as time evolves it oscillates around zero alternating between positive and negative values, that is between super and sub-Poissonian statistics this happens in the same temporal region where the exchange of excitations between the field and the atom is most important. After some time it oscillates above zero with a small amplitude (when the probability to find the atom in its excited state is constant) and remains with a super-Poissonian statistics until the revival time (see figure \ref{fig_3}) when the oscillations around zero repeat themselves. 

\section{Conclusions}
In this work we have presented an approximate method to construct the time evolution operator for  a hybrid system composed by a forced optomechamical oscillator and a two-level atom inside the cavity, the atom interacts only with the cavity field by means of a Jaynes-Cummings interaction. In order to solve the problem we split the Hamiltonian as the sum of a forced optomechanical Hamiltonian and that of the free atom with the JC interaction. The time evolution operator for the forced optomechanical Hamiltonian is approximated as a product of exponentials \cite{optoforzado} and it is then used to take the JC interaction into a generalized interaction picture. As a result we obtained cumbersome expressions for the transformed operators which we approximated by neglecting terms of the order $G/\omega_m$ and $(G/\omega_m)^2$ as compared with the cavity frequency $\omega_c$. Within this approximation the interaction picture Hamiltonian becomes that of a free two level atom and a displaced JC interaction whose exact  time evolution operator we constructed using the Wei-Norman Theorem. Once we have the full time evolution operator we can obtain the average value of any observable, as an example we evaluated the temporal evolution of the average value of the photon and phonon number operators, the probability to find the atom in its excited state and the Mandel parameter for the cavity field. We used as initial state $|\Psi(0)\rangle = |\alpha\rangle\otimes |e\rangle \otimes |\Gamma\rangle$ where $|\alpha\rangle$ is the  ket corresponding to the cavity field in a coherent state $\alpha$, $|e\rangle$ is that corresponding to the atom in its excited state and $|\Gamma\rangle$ is  the ket for the mechanical oscillator in a coherent state $\Gamma$. The average number of photons is a function of the pumping amplitude $\Omega$ and the pump frequency $\omega_L$,  when $\omega_L\simeq \omega_c$ there is a periodic growth in the number of photons and the amplitude of this growth is proportional to $\Omega$. The average number of phonons is a periodic function of time which depends also on the optomechanical coupling  $G/\omega_m$ and on the number of photons present in the cavity. For red detuning there is a power flow from the mechanical mode to the optical mode and the cooling of the mechanical mode is more important as the number of photons increases. Since the evolution of the phonon number is periodic one can select an interaction time such that the number of phonons be at a minimum. We also evaluated the Mandel parameter for the cavity field and we found that it alternates between sub-Poissonian and super-Poissonian statistics in the region of time where there is an important exchange of excitations between the atom and the cavity field. We stress the fact that our approximations are done in the interaction Hamiltonian where we have neglected terms proportional to $G/\omega_m$ and $(G/\omega_m)^2$ with respect to the cavity frequency $\omega_c$. The excellent agreement between the analytic and the numerical results 
obtained using the full Hamiltonian as given in \eqref{eq:hamiltonian} indicate the validity of our approximations.

\appendix
\section{The Wei–Norman approach}\label{apA}
\setcounter{equation}{0}
\renewcommand{\theequation}{A{\arabic{equation}}}
Here we describe the method we used to obtain the time evolution operator for the forced optomechanical system.  The first thing to notice is that the set of operators appearing in $\hat H_{opt}$ is closed under commutation
\begin{center}
	\begin{tabular}{|c|c|c|c|c|c|}
		\hline
		&$\hat n$&$\hat N$&$\hat n\hat b$&$\hat n \hat b^{\dagger}$&$\hat
		n^2$\\
		\hline
		$\hat n$&0&0&0&0&0 \\
		\hline
		$\hat N$&0&0&$-\hat n\hat b$&$\hat n\hat b^{\dagger}$&0 \\
		\hline
		$\hat n\hat b$&0&$\hat n\hat b$&0&$\hat n^2$&0 \\
		\hline
		$\hat n\hat b^{\dagger}$&0&$-\hat n\hat b^{\dagger}$&$-\hat
		n^2$&0&0 \\
		\hline
		$\hat n^2$&0&0&0&0&0\\
		\hline
	\end{tabular}
\end{center}
In this table, we had to incorporate the operator $\hat n^2$ that arises from the commutator between $\hat n\hat b$ and $\hat n \hat b^{\dagger}$.  The time evolution operator corresponding to the Hamiltonian $\hat H_{opt}$ can then be written {\em exactly} as a product of exponentials \cite{mancini,Wei_Norman},

\begin{equation}\label{eq:opevol}
\hat U_{opt}(t) = e^{\alpha_1 \hat n}e^{\alpha_2 \hat N}e^{(\alpha_3+|\alpha_4|^2/2) \hat n^2}\hat{D}_{\hat{b}}(\alpha_4\hat{n}).
\end{equation}
The time-dependent functions $\alpha_i$ are obtained after substitution of Eq.~\ref{eq:opevol} into Schr\"odinger's equation. As a result we get:
\begin{equation}
\begin{split}
\alpha_1& = -i \omega_c t,\\
\alpha_2& = -i \omega_m t,\\
\alpha_3& = -\left(\frac{G}{\omega_m}\right)^2\left[-i \omega_m t+\left(1-e^{-i \omega_m t}\right)\right],\\ 
\alpha_4& = -\frac{G}{\omega_m}(1-e^{i\omega_m t}).
\end{split}
\end{equation}
Once we know the exact time evolution operator for the optomechanical system, we transform the forcing term to obtain an interaction picture Hamiltonian 
\begin{equation}
\hat H_I^{(0)} =\hbar  \Omega \cos(\omega_L t) \left[\hat U_{opt}(t)^{\dagger} (\hat a+\hat a^{\dagger}) \hat U_{opt}(t)\right],
\end{equation}
applying the transformation we obtain:
\begin{equation}
\begin{split}
\hat U_{opt}^{\dagger}\hat a \hat U_{opt}&=
 e^{i E(t)(2\hat n+1)}e^{i F(t)[\hat b^{\dagger}e^{i\frac{\omega_m}{2}t}+\hat b e^{-i\frac{\omega_m}{2} t}]}\hat a e^{-i\omega_c t},\\
\hat U_{opt}^{\dagger}\hat a^{\dagger} \hat U_{opt}&= \hat a^{\dagger} e^{i\omega_c t}e^{-i F(t)[\hat b^{\dagger}e^{i\frac{\omega_m}{2}t}+\hat b e^{-i\frac{\omega_m}{2}t}]} e^{-i E(t)(2\hat n+1)},
\end{split}
\end{equation}
where 
\begin{equation}
\begin{split}
F(t)& = 2\left(\frac{G}{\omega_m}\right)\sin\left(\frac{\omega_m}{2} t\right),\\ 
E(t)&= \left(\frac{G}{\omega_m}\right)^2 (\omega_m t - \sin(\omega_m t)).
\end{split}
\end{equation}
Notice the presence of the operators in the exponentials. However, the factor $G/\omega_m \ll 1$ \cite{Ventura} so that we make the approximation
\begin{equation}
\hat U_{opt}^{\dagger}\hat a \hat U_{opt}\simeq \hat a e^{-i\omega_c t}, \ \ \  \hat U_{opt}^{\dagger}\hat a^{\dagger} \hat U_{opt}\simeq \hat a^{\dagger} e^{i\omega_c t},
\end{equation}
and, in this approximation, we get the interaction Hamiltonian
\begin{equation}\label{eq:hiap}
\tilde H_I^{(0)} = \hbar \Omega \cos(\omega_L t) \left( \hat a e^{-i\omega_c t}+\hat a^{\dagger} e^{i\omega_c t}\right), 
\end{equation}
where we have used $\tilde H_I^{(0)}$ to denote the approximate interaction Hamiltonian. The corresponding time evolution operator can be written as a product of exponentials
\begin{equation}
\hat U_I^{(0)} = e^{\beta \hat a^{\dagger}} e^{\gamma \hat a} e^{\delta},
\end{equation}
with:
\begin{equation}
\begin{split}
\dot \beta &= -i\Omega \cos(\omega_L t) e^{i\omega_c t}, \\
\dot \gamma&= -i\Omega \cos(\omega_L t) e^{-i\omega_c t}, \\
\dot \delta&= \beta \dot \gamma,
\end{split}
\end{equation}
and initial conditions $\beta(0)=\gamma(0)=\delta(0)=0.$ We see that $\beta = -\gamma^{*}$ so that we can write
\begin{equation}
\begin{split}
\hat U_I^{(0)}&= e^{\delta} e^{\frac{1}{2}|\beta|^2} e^{\beta \hat a^{\dagger}-\beta^{*}\hat a} \\
&= e^{\delta +\frac{1}{2}|\beta|^2}\hat D_{\hat a}(\beta). 
\end{split}
\end{equation}
Finally, taking into account the above relationships and equation \eqref{eq:hopt}, the approximate evolution operator of the forced optomechanical system is \cite{optoforzado}
\begin{equation}
\begin{split}
\hat U_0&= \hat U_{opt} \hat U_I^{(0)}\\
&=  e^{\delta +\frac{1}{2}|\beta|^2} e^{\alpha_1 \hat n}e^{\alpha_2 \hat N}e^{(\alpha_3+|\alpha_4|^2/2) \hat n^2}\hat{D}_{\hat{b}}(\alpha_4\hat{n}) \hat D_{\hat a}(\beta).
\end{split}
\end{equation}

\section*{Acknowledgments}
We thank Reyes Garcia for the maintenance of our computers. We acknowledge partial support from Direcci\'on General de Asuntos del Personal Acad\'emico, Universidad Nacional Aut\'onoma de M\'exico (DGAPA UNAM) through project PAPIIT IN 1111119 and I. Ramos-Prieto acknowledges  postdoctoral support from DGAPA UNAM.

\end{document}